\documentclass{PoS}
\usepackage{cite}
\usepackage{graphicx}
 \usepackage{amsmath}
\usepackage{amssymb} 
\usepackage{epsfig}
\usepackage{color}
\usepackage{latexsym}
\usepackage{amsmath}
\usepackage{amssymb}
\usepackage{dsfont}
\usepackage{verbatim}

\title{Lattice spectroscopy (focus on exotics)}

\ShortTitle{Lattice spectroscopy }

\author{\speaker{Sasa Prelovsek} \\
        Faculty of Mathematics and Physics, University of Ljubljana, Slovenia\\
Jozef Stefan Institute, 1000 Ljubljana, Slovenia\\
Instit\"ut f\"ur Theoretische Physik, Universit\"at Regensburg, D-93040 Regensburg, Germany
        E-mail: \email{sasa.prelovsek@ijs.si}}

\abstract{Recent lattice QCD results on the hadron spectroscopy with beauty and charm quarks are reviewed.  The focus of the review is exotic hadrons, while non-lattice approaches and conventional hadrons are reported as well.  We  discuss  the recently discovered $\Omega_c^*$, a charmonium state with spin three, $P_c$ pentaquarks, $B_c(2S)$ and $\Lambda_b^*$,  the long-standing challenges for theory $Z_c$, $Z_b$ and $X(3872)$, and we review predictions for yet undiscovered states $bb\bar q\bar q$, $\bar bb\bar bb$, highly excited and hybrid $\bar bb$,  baryons with bottom quarks and still-missing $B_{s}$ mesons.  
 }

\FullConference{18th International Conference on B-Physics at Frontier Machines - Beauty2019 -\\
		29 September / 4 October, 2019\\
		Ljubljana, Slovenia}

\begin{document}

\section{Introduction}

The spectroscopy of hadrons  that contain beauty and charm quarks is reviewed. The presentation is divided to a part   on the recently discovered states, the long-standing challenges for theory and predictions of yet-undiscovered hadrons. Many more details can be found in valuable recent reviews on various aspects of this subject  \cite{Brambilla:2019esw,Kou:2018nap,Ali:2019roi,Guo:2017jvc,Liu:2019zoy, 
Chen:2016qju,Esposito:2016noz,Ali:2017jda}.

\section{Hadrons from lattice QCD: resonances and (shallow) bound states}

The physics information on a hadron (below, near or above threshold) is commonly extracted from the energies $E_n$ of QCD eigenstates $|n\rangle$ on a finite and discretized lattice. The eigen-energies $E_n$ are determined from the time-dependence of  the correlation functions 
$\langle O_i (t) O_j^\dagger (0)\rangle=$

\noindent
 $=\sum_{n} \langle O_i|n\rangle  ~e^{-E_n t} \langle n|O^\dagger_j\rangle~,$
  where operators $O$ create/annihilate the hadron system with a given quantum number of interest.  

The energy of a strongly-stable hadron with zero momentum directly gives a hadron mass if this hadron is significantly below thresholds. For example, $E_1(p=0,J^P=0^-)=m_B$ for a ground state hadron with flavor $b\bar u$. 

 \begin{figure}[htb]
\centering
\includegraphics[width=0.4\textwidth,clip]{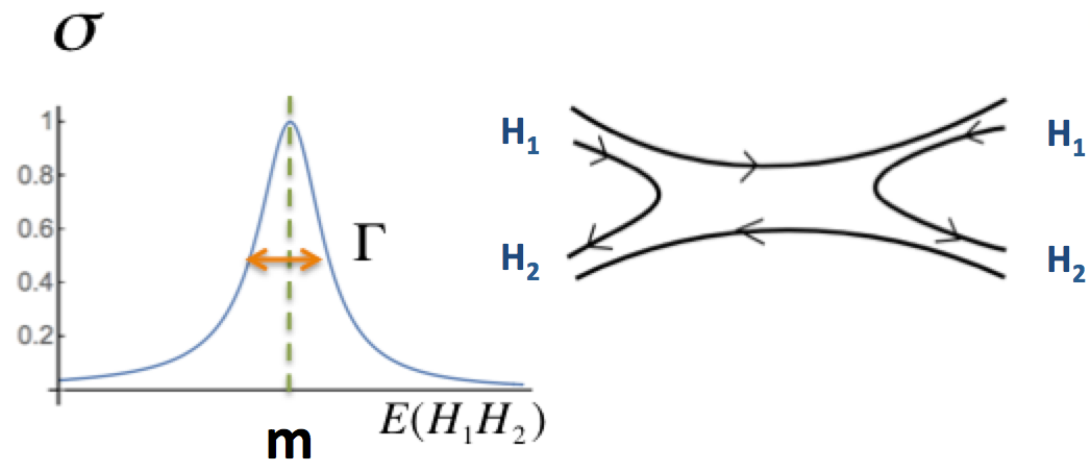}  $\quad$
\includegraphics[width=0.1\textwidth,clip]{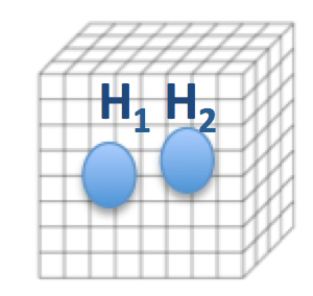}  $\quad$
\includegraphics[width=0.35\textwidth,clip]{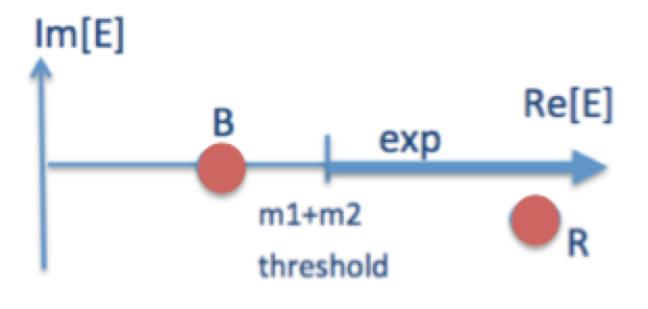}  
\caption{Hadronic resonances ($R$) and bound states ($B$) from scattering of two hadrons $H_1H_2$ on the lattice.   The right plot shows location of the poles in the scattering matrix $T(E)$ in the complex energy plane. }
\label{fig:1}
\end{figure}

 In the energy region near or above threshold, the masses of bound-states and resonances have to be inferred from the scattering of two hadrons $H_1H_2$, which is encoded in the    scattering matrix $T(E)$  (Fig. \ref{fig:1}).    The simplest example is a one-channel (elastic) scattering  in partial wave $l$, where   the scattering matrix $T(E)$ has size $1\times 1$.     L\"uscher has shown that the energy  $E$ of two-hadron eigenstate in finite volume $L$  (Fig. \ref{fig:1}) gives the scattering matrix $T(E)$  at that energy in infinite volume  \cite{Luscher:1991cf}.  This relation and 
 its generalizations are at the core of extracting rigorous  information about the scattering from the lattice simulations.  It leads  to $T(E)$  for real $E$  above and somewhat below $H_1H_2$ threshold.  The  resulting $T(E)$   provides the masses of resonances and bound states:  
  \begin{itemize}
\item  In the vicinity of a {\it hadronic resonance} with mass $m_R$ and width $\Gamma$, the   scattering matrix has a Breit-Wigner-type shape   $T(p)=\frac{-\sqrt{s}~ \Gamma(p)}{s-m_R^2+i \sqrt{s}\, \Gamma(p)}$ with $
\Gamma(p)=g^2\,\frac{p^{2l+1}}{s}$. One can extract $m_R$ and $\Gamma$ from the position of the peak and the width in the cross section $\sigma \propto |T(p)|^2$. Or one can  continue $T(E)$ to complex $E$: then the position of the pole in $T(E)$ at $E=m_R-i\frac{1}{2}\Gamma$ 
renders the resonance parameters (Fig. \ref{fig:1}).  
\item The {\it bound state (B)} in $H_1 H_2$ scattering is realized when $T(E)$ has a pole for real energy below threshold: $T(E=m_B)=\infty$ (Fig. \ref{fig:1}).  
This can be easily understood from the propagator of the bound-state $1/(p^2-m_B^2)$ in s-channel  $H_1H_2$ scattering, which is infinite for $p^2=E^2=m_B^2$ in the center-of-momentum frame. The state is referred to the bound state if the pole occurs for positive imaginary momenta $p=i|p|$ and a {\it virtual bound state} if it occurs for $p=-i|p|$, where $p$   denotes the magnitude of the 3-momentum in the center-of-mass frame. 

\end{itemize}

\section{Hadrons that were recently discovered or confirmed  in experiment}

\underline {Doubly-charmed baryon $\Sigma_{cc}$: $ccq$}

\begin{figure}[h!]
\centering
\includegraphics[width=0.35\textwidth,clip]{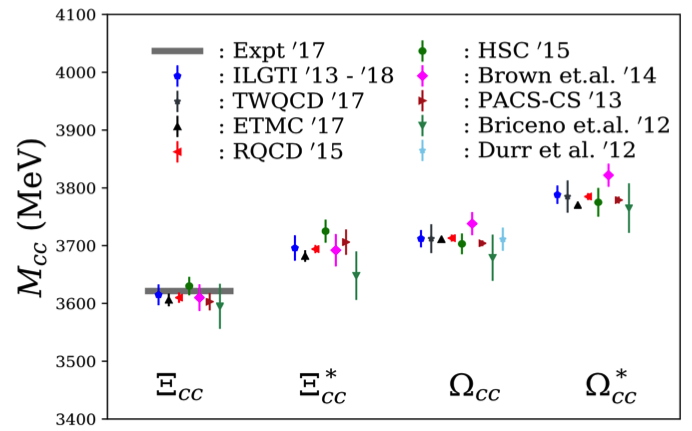}  $\quad$
\includegraphics[width=0.35\textwidth,clip]{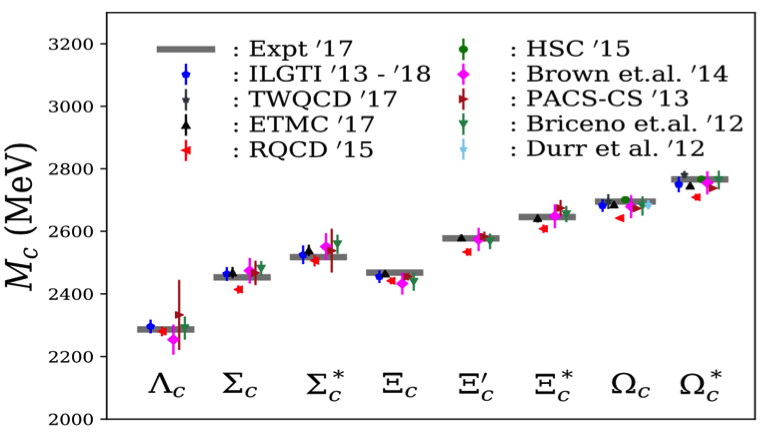}  
\caption{ Masses of singly and doubly charmed baryons from lattice QCD (points) and experiment (gray lines). Compilation taken from \cite{Padmanath:2019wid}.    }
\label{fig:2}
\end{figure}

Various lattice QCD {\it pre}dictions for the mass of a doubly charmed baryon $\Sigma_{cc}$ are collected in Fig. \ref{fig:2} and they agree well with the LHCb discovery \cite{Aaij:2017ueg}. Note that the mass splitting of the $\Sigma_{cc}$ isospin  partners is determined to be only  around  $2~$MeV by an impressive lattice QCD study by BMW \cite{Borsanyi:2014jba}. This makes it unlikely that the SELEX experiment  found the isospin partner at a much lower mass.   Other lattice QCD predictions and postdictions  for singly and doubly charmed baryons are also collected in Fig. \ref{fig:2}. All these follow directly from $E_1(p=0)=m$ since these states are below thresholds. 
 
 \begin{figure}[h!]
\centering
\includegraphics[width=0.4\textwidth,clip]{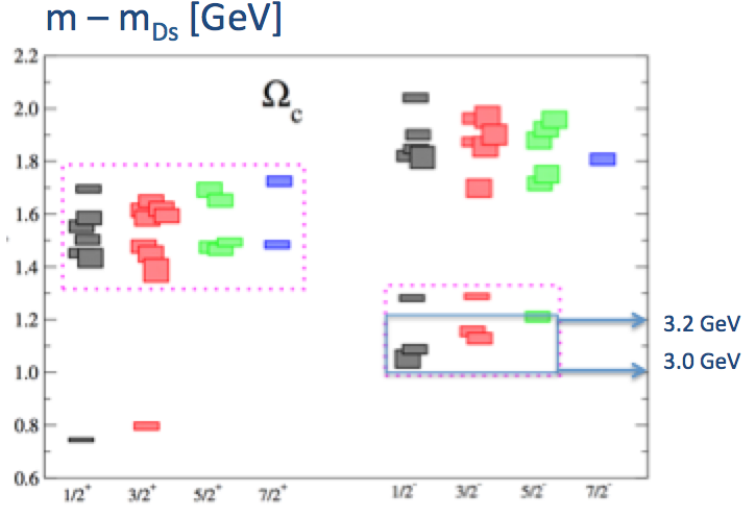}   
\includegraphics[width=0.25\textwidth,clip]{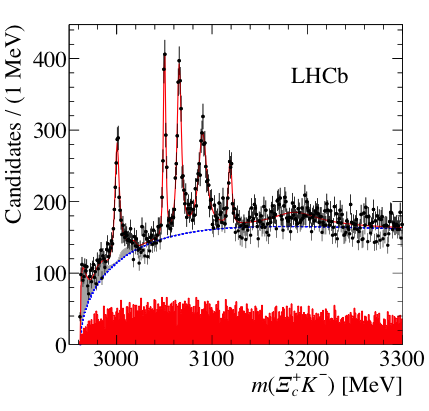} 
\includegraphics[width=0.33\textwidth,clip]{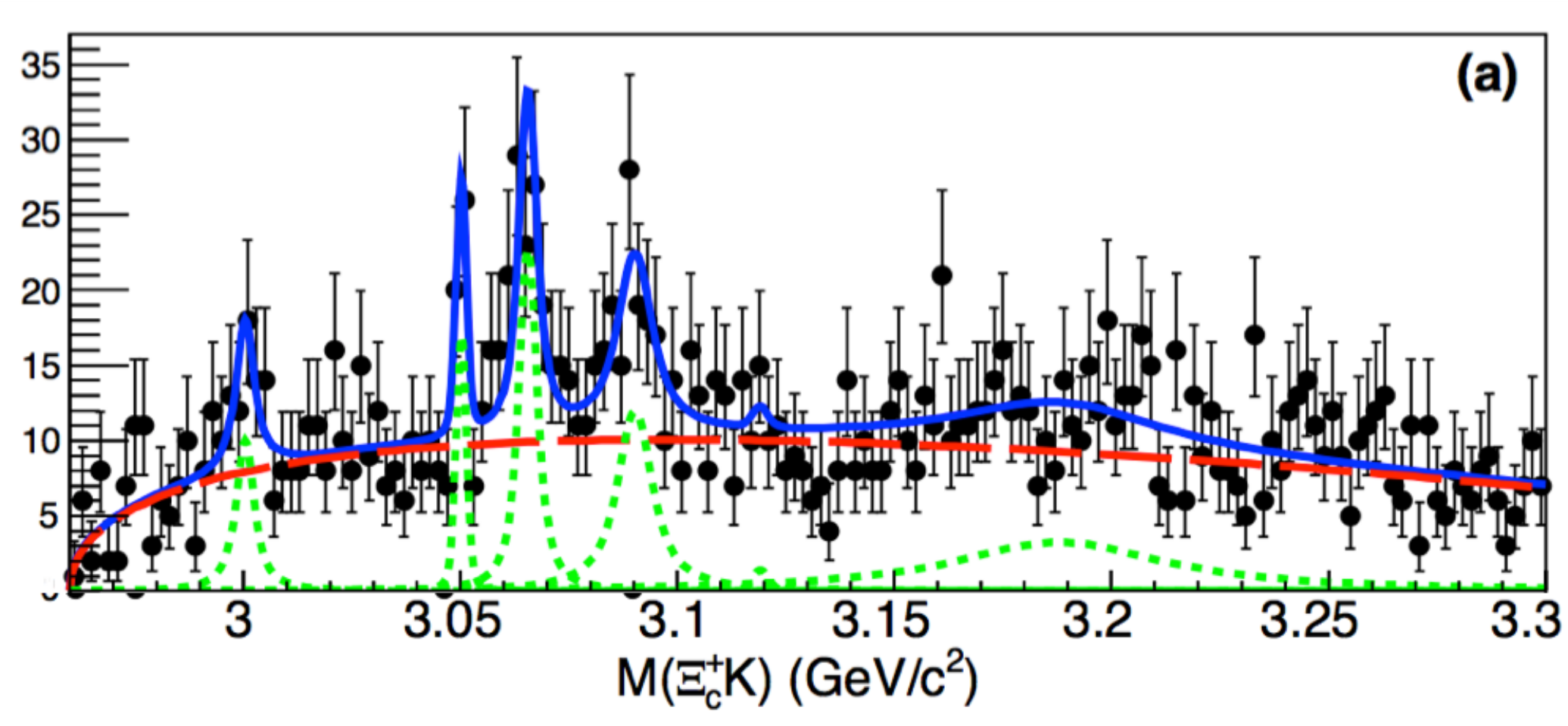}  
\caption{ Left pane:   $\Omega_c$ states with quark content $css$ and various $J^P$ as predicted from lattice in 2013 \cite{Padmanath:2013bla}: every box denotes a predicted $\Omega_c$ meson.  Right two panes: the newly observed $\Omega_c$ states by LHCb  \cite{Aaij:2017nav} and Belle  \cite{Yelton:2017qxg} in 2017.  }
\label{fig:3}
\end{figure}

 \underline {Five excited  $\Omega_{c}^*$: $css$}
 
 LHCb \cite{Aaij:2017nav}  discovered five excited $\Omega_{c}^*$ and Belle recently confirmed four of them \cite{Yelton:2017qxg}. All these states were predicted by lattice QCD already in 2013 \cite{Padmanath:2013bla}: there are indeed five states in the energy region between $3.0-3.2$ GeV in Fig \ref{fig:3}a. This lattice study predicted their quantum numbers to be $1/2^-,~1/2^-,~3/2^-,3/2^-,~5/2^-$ respectively. Each box in Fig \ref{fig:3}a presents a predicted $\Omega_c$ baryon, where predictions of higher-lying states ignore their strong decays.  A more detailed and recent lattice simulation \cite{Padmanath:2017lng} confirms previous predictions. The discovered $\Omega_{c}^*$ are believed to be conventional $css$ states. \\

 \underline { Charmonia with  $J^{PC}=3^{--}$ and $1^{--}$: $\bar cc$}
 
\begin{figure}[h!]
\centering
\includegraphics[width=0.2\textwidth,clip]{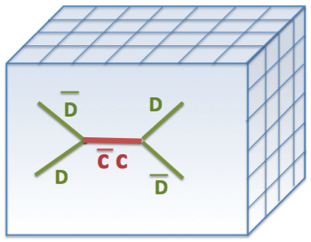}    $\qquad$
\includegraphics[width=0.33\textwidth,clip]{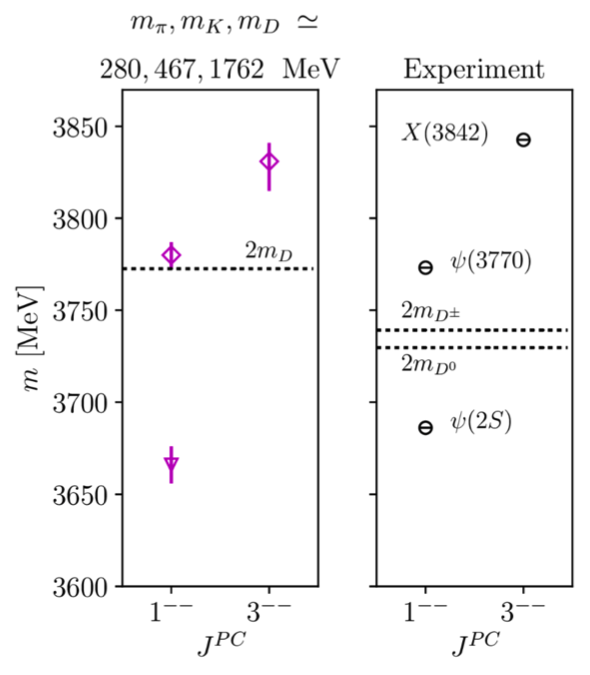}   $\qquad$
\includegraphics[width=0.22\textwidth,clip]{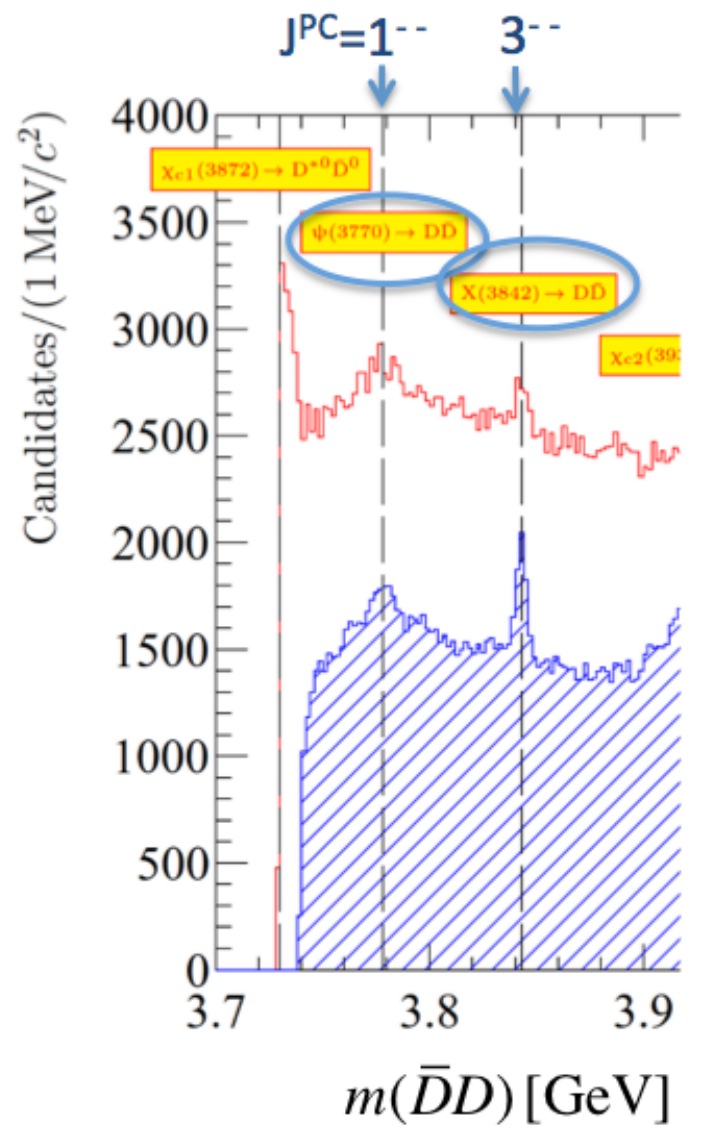}  
\caption{ The resonance masses of charmonia with $J^{PC}=3^{--}$ and $1^{--}$, extracted from lattice \cite{Piemonte:2019cbi} and compared to experiment \cite{Aaij:2019evc}. }
\label{fig:4}
\end{figure}

 Charmonium resonances with  $J^{PC}=3^{--}$ and $1^{--}$    have been extracted from lattice QCD by simulating $D\bar D$ scattering in partial waves $l=3,~1$  \cite{Piemonte:2019cbi}, as described in Section 2. The resonance with $J^{PC}=3^{--}$ was found at a mass that is in agreement with the very recent LHCb  discovery of a first charmonium $X(3842)$ with spin three \cite{Aaij:2019evc} (Fig. \ref{fig:4}). The width of this resonance is too small to resolve from a lattice simulation. The lowest vector resonance $\psi(3770)$ above $D\bar D$ threshold was  found on the lattice at the mass close to experimental mass and in this case the width was also extracted. The resulting coupling $g^{lat}=16.0{+2.1\atop -0.2}$ that parametrizes the width via $\Gamma=g^2p^3/(6\pi s)$ is consistent with $g^{exp}=18.7\pm 0.9$.  These charmonia are conventional $\bar cc$ states, dominated by  $n~^{2s+1}l_j=1~^3D_3$ and $1~^3D_1$ according to the  quark-model assignment.\\

 \underline {  $P_c$ pentaquarks: $\bar ccuud$}

 The LHCb experiment confirmed/discovered  three $P_c$ pentaquarks  near $\Sigma_c^+\bar D^{(*)}$ thresholds in 2019  \cite{Aaij:2019vzc} and all of them were found in the $J/\psi~p$ decay channel. 
Rigorous theoretical treatment of $P_c$ presents an enormous challenge because these hadrons can strongly decay into several channels. Most success was achieved by simplified phenomenological treatment which describes  the lower $P_c$ as  a $\Sigma_c^+\bar D$  molecule with $J^P=1/2^-$ in s-wave, and the higher pair of $P_c$ as   $\Sigma_c^+\bar D^*$ which can combine to $J^P=1/2^-$ or $3/2^-$ in s-wave \cite{Wu:2010jy,Wu:2012md,Yang:2011wz,Wang:2011rga}. The pion exchange in $t$-channel of    $\Sigma_c^+\bar D$ is forbidden by parity \cite{Karliner:2015ina} , so $t$-channel exchange of vector mesons $\rho,~\omega,~\phi$  provides the crucial interaction in these models.  Note that some of these models predict also states that have not been observed by experiment. 

The only available lattice QCD simulation that reaches the energies of $P_c$ resonances  addressed the simplified question: do the $P_c$ resonances appear in one-channel $J/\psi p \to P_c\to J/\psi p$ scattering where this channel is decoupled from other channels? The answer to this question from a lattice simulation \cite{Skerbis:2018lew} is: No. This indicates that the coupling of $J/\psi p$ channel with other two-hadron channels is responsible for the existence of $P_c$ in experiment. This is in line with LHCb results, which suggests that the coupling of $J/\psi p$ to the open-channel $\Sigma_c^+\bar D^{(*)}$ is essential.  \\

 \underline {  $B_c(2S)$ and $B_c^*(2S)$ : $b \bar c$}
 
\begin{figure}[h!]
\centering
\includegraphics[width=0.6\textwidth,clip]{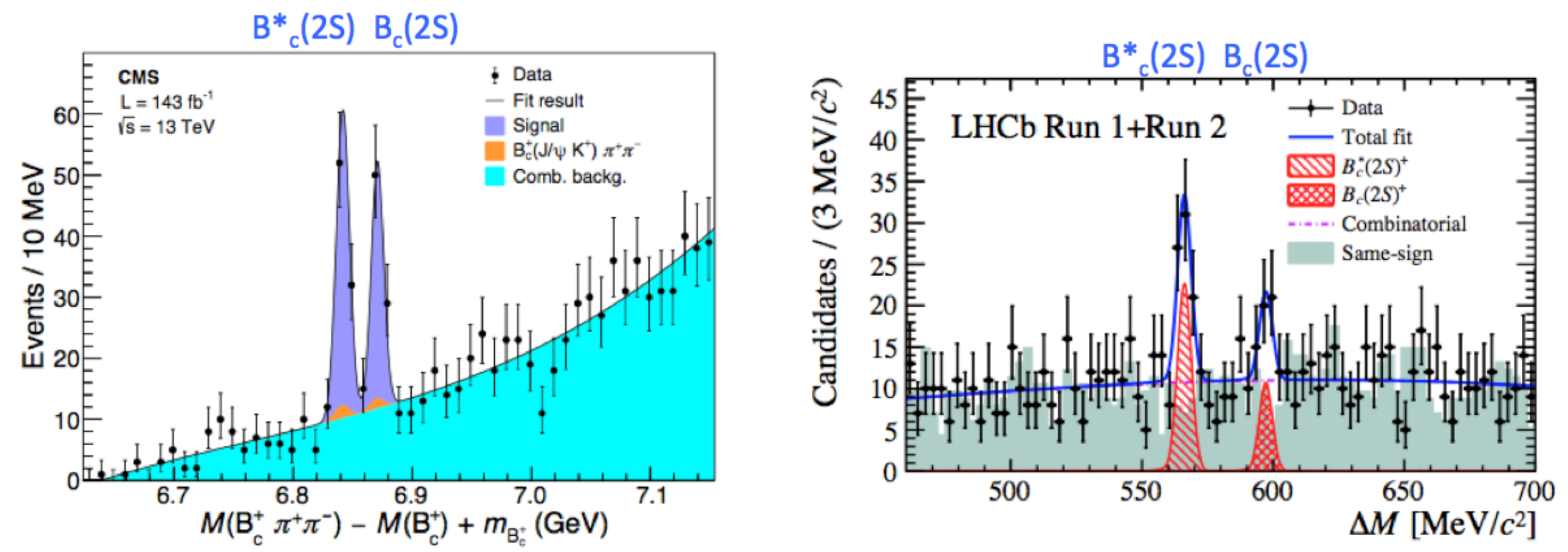}    $\quad$
\includegraphics[width=0.33\textwidth,clip]{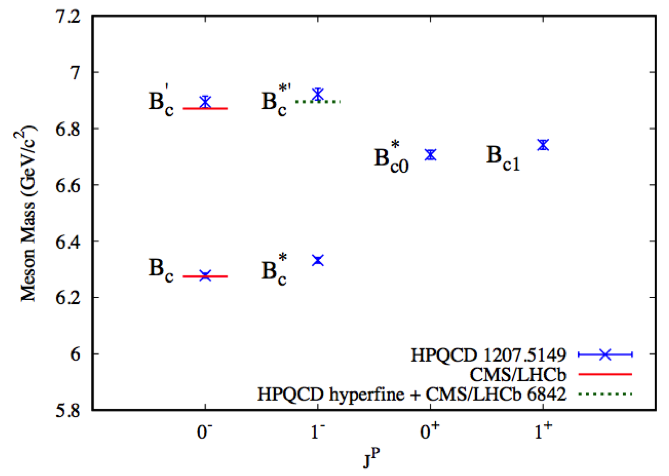}    
\caption{  The  $B_c(2S)$ and $B_c^*(2S)$ peaks from CMS \cite{Sirunyan:2019osb} and  LHCb \cite{Aaij:2019ldo}, together with predicted masses from lattice QCD (blue crosses in right pane) \cite{Dowdall:2012ab}.}
\label{fig:5}
\end{figure}

 The masses of discovered $B_c(2S)$ and $B_c^*(2S)$ by CMS \cite{Sirunyan:2019osb} and  LHCb \cite{Aaij:2019ldo} agree with the lattice QCD predictions by HPQCD \cite{Dowdall:2012ab,Lytle:2018ugs} in Fig. \ref{fig:5}. The lattice value of the hyperfine splitting $m[B_c^*]-m[B_c]$  is used to convert the the experimental $\Delta M^{exp}\equiv m[B_c(2S)]-m[B_c^*(2S)]=\{m[B_c^*]-m[B_c]\}-\{m[B_c^*(2S)]-m[B_c(2S)]\}$ to $m[B_c^*(2S)]$ (green dashed line), since the photon in $B_c^*\to B_c\gamma$ is undetected.   \\
 
 \underline {  New excited $\Lambda_b^0$:  $bdu$} 
 
 The quark model predictions (see Table VI of \cite{Chen:2014nyo}) indicate that the two newly discovered $\Lambda_b^0$  \cite{Aaij:2019amv} are $J^P=3/2^+$ and $5/2^+$ states. There are no available  lattice QCD results for these states, while  the spectrum  for the charmed partners  was calculated in  \cite{Padmanath:2013bla}. \\

  \begin{figure}[h!]
\centering
\includegraphics[width=0.23\textwidth,clip]{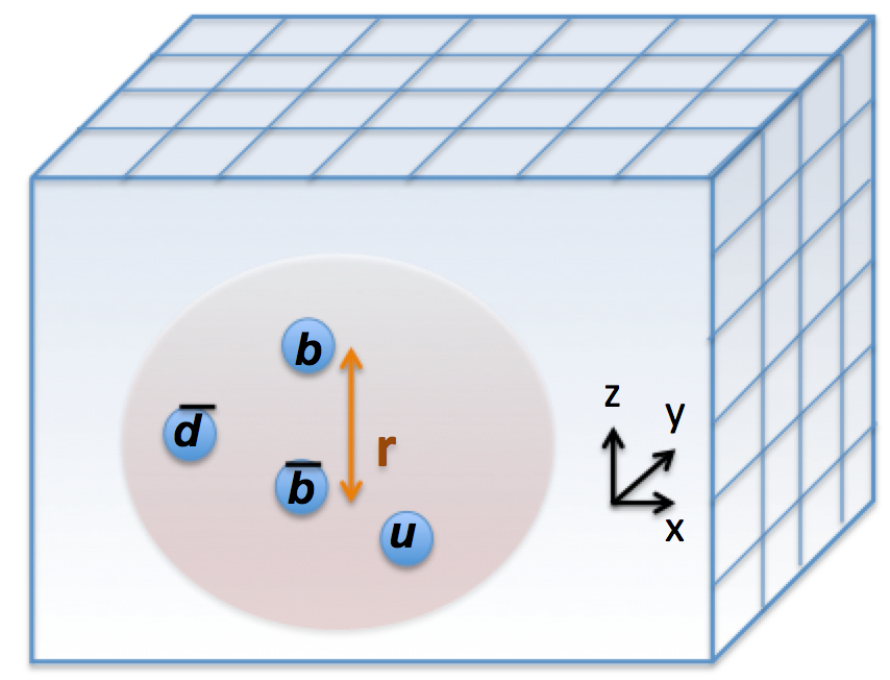}    $\quad$
\includegraphics[width=0.23\textwidth,clip]{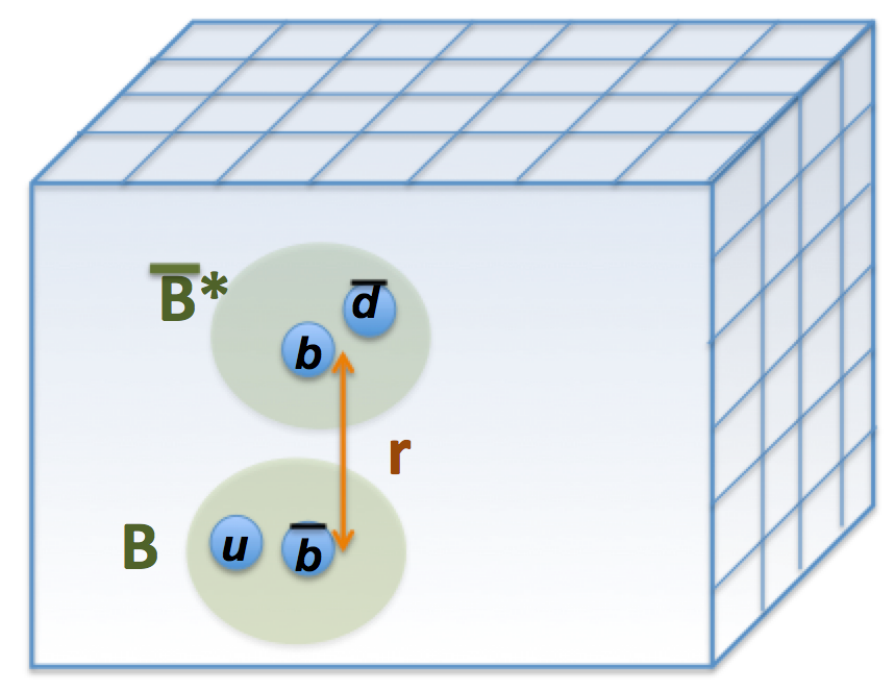}  $\qquad$  
\includegraphics[width=0.28\textwidth,clip]{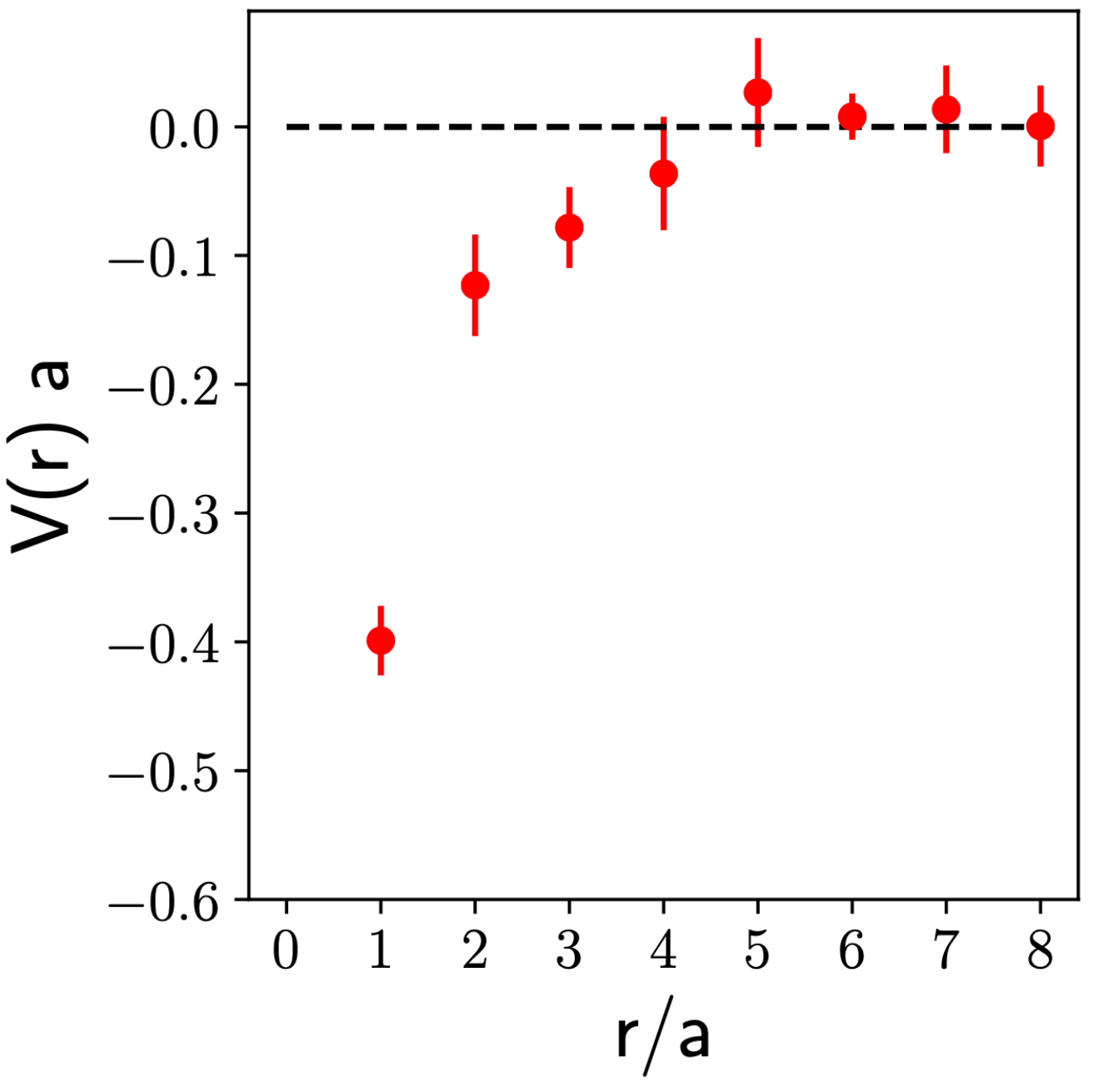} 
\caption{  Simulation of  $\bar bb\bar du$ channel with static $b\bar b $ pair as a function of their distance $r$. The extracted potential between $B$ and $\bar B^*$ is shown on the right \cite{Prelovsek:2019ywc}. }
\label{fig:7}
\end{figure}

 \section{Hadrons that are long-standing challenges for theory}

 \underline {$Z_b^+$: $\bar bb\bar du$}
 
 Belle discovered two exotic $Z_b^+$ states with $J^P=1^+$ and quark content $\bar bb\bar du$ near $B^{(*)}\bar B^*$ threshold in five decay channels \cite{Belle:2011aa,Garmash:2015rfd}.  The lattice simulations \cite{Prelovsek:2019ywc,Peters:2016wjm} study this channel by exploring the energy of the system as a function of the distance $r$ between static $b$ and $\bar b$ in Fig. \ref{fig:7} (for the case of total spin of $\bar bb$ equal to $1$). The important finding is that the energy of  the $B\bar B^*$ eigenstate is significantly lower than $m_B+m_{B^*}$ at small $r$. This gives evidence for significantly attractive interaction $V(r)$ between $B$ and $\bar B^*$ at small distances  \cite{Prelovsek:2019ywc} (Fig. \ref{fig:7}). The lattice potential   leads to a  bound state  with the mass with the mass $M-m_B-m_{B^*}=- 48^{~+41}_{~-108} ~\mathrm{MeV}$. 
The significant uncertainty of the binding energy captures the statistical errors  
 as well as  various choices for  parametrizing the potential.   
 The presence of this bound-state pole renders a peak in the $B\bar B^*$ rate above threshold 
 if the bound state lies closely below threshold. This is illustrated in  Fig.  \ref{fig:8}, which  shows this rate  for three fits that are all consistent with our lattice potential.   The bound state with a small binding energy $\simeq 8~$MeV leads to a peak in the $B\bar B^*$ rate  above threshold.    Its shape    resembles the $Z_b(10610)$ peak in the $B\bar B^*$ rate observed by Belle (Fig. 2 of \cite{Garmash:2015rfd}).   These exploratory lattice studies suggest that attraction between $B$ and $\bar B^*$ is responsible for the existence of $Z_b$ exotic hadrons. \\

 \begin{figure}[h!]
\centering
\includegraphics[width=0.29\textwidth,clip]{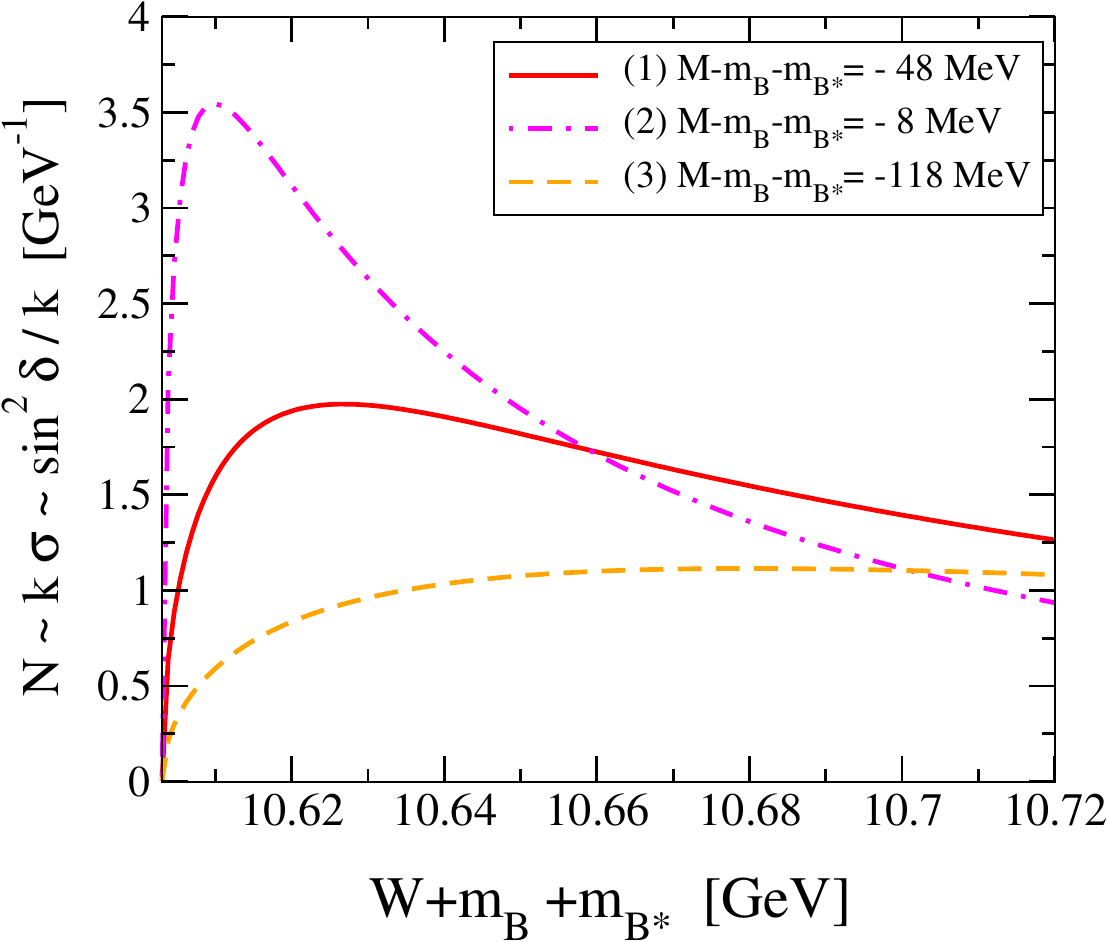}    $\qquad$
\includegraphics[width=0.38\textwidth,clip]{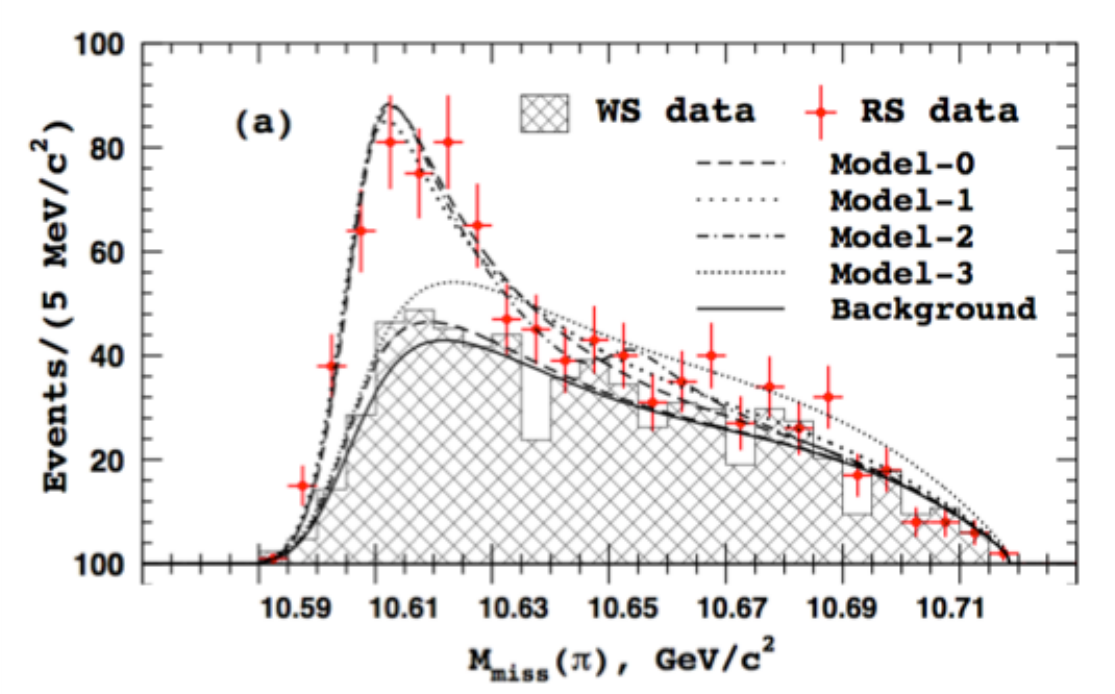}    
\caption{ The $Z_b\to B\bar B^*$ rate $N$  from lattice QCD study \cite{Prelovsek:2019ywc} (left) and from Belle \cite{Garmash:2015rfd} (right). Three  choices of fits   consistent with  the lattice potential   are shown }
\label{fig:8}
\end{figure}

 \underline {$Z_c^+$: $\bar cc\bar du$}
 
 The consensus on the nature of exotic $Z_c^+(3900)$ discovered by BESIII \cite{Ablikim:2013xfr}  and Belle \cite{Liu:2013dau} has not been achieved yet. The reanalysis of experimental data is compatible with several scenaria \cite{Pilloni:2016obd}. The lattice QCD study \cite{Ikeda:2016zwx} based on the less-rigorous  HALQCD method suggests that the coupling between $D\bar D^*$ and $J/\psi \pi$ channels is responsible for the existence of $Z_c(3900)$: see Fig. \ref{fig:8}. The lattice study \cite{Chen:2019iux} based on the more-rigorous L\"uscher's method has unfortunately not found evidence for $Z_c(3900)$ (yet), confirming a previous lattice result \cite{Prelovsek:2014swa}.
 
    \begin{figure}[h!]
\centering
\includegraphics[width=0.38\textwidth,clip]{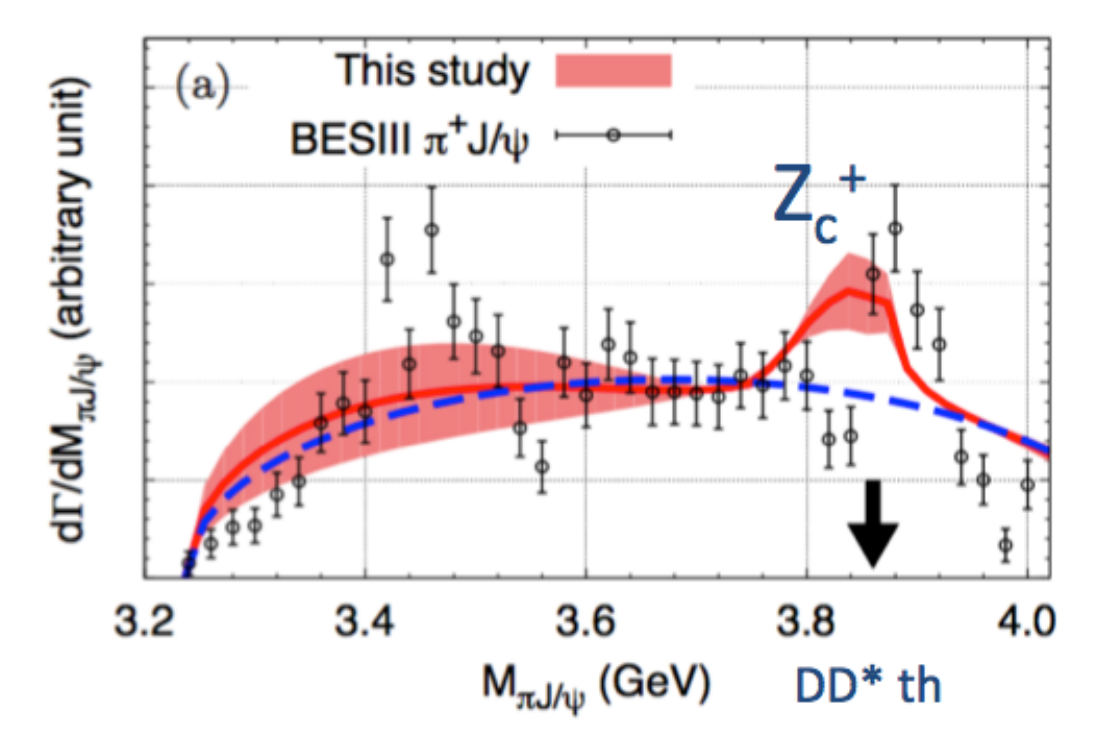}    $\qquad$
\includegraphics[width=0.35\textwidth,clip]{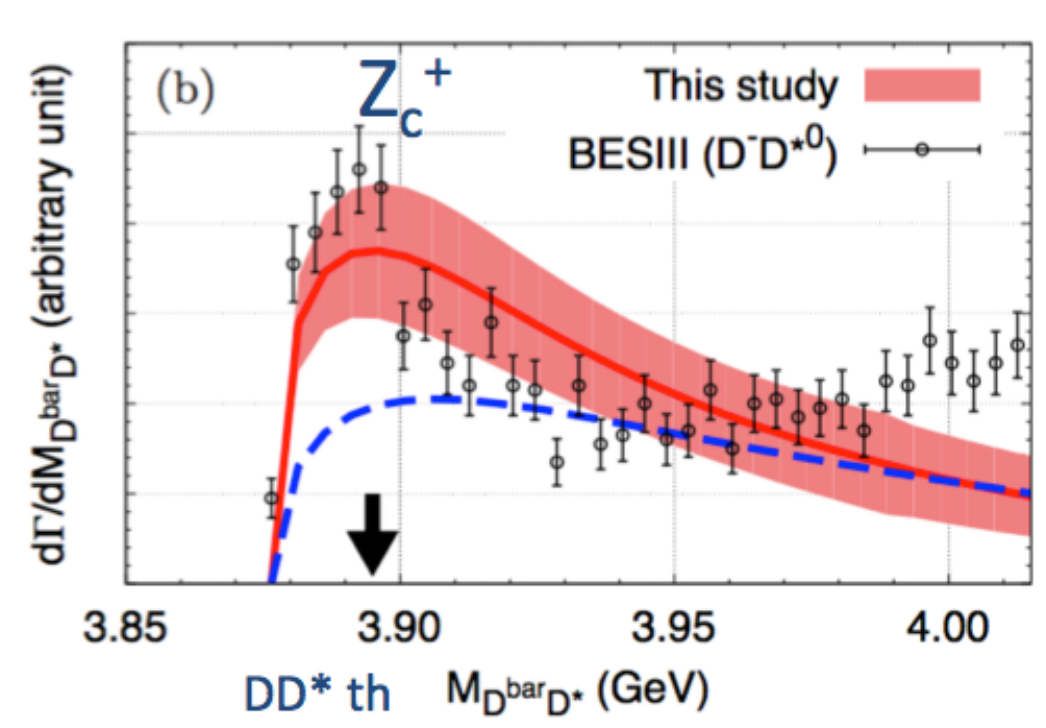}    
\caption{  Red line: the rate obtained from lattice shows a $Z_c$ peak \cite{Ikeda:2016zwx}. Blue dashed line:  the rate obtained from lattice if the coupling $D\bar D^*$-$J/\psi \pi$ is enforced to zero does not have $Z_c$ peak  \cite{Ikeda:2016zwx}. Black points: the rate by BESIII  \cite{Ablikim:2013xfr}.}
\label{fig:8}
\end{figure}

\underline {$\chi_{c1}(2P)$ aka $X(3872)$: $\bar cc$ + $\bar cq\bar qc$}

\begin{figure}[h!]
\centering
\includegraphics[width=0.2\textwidth,clip]{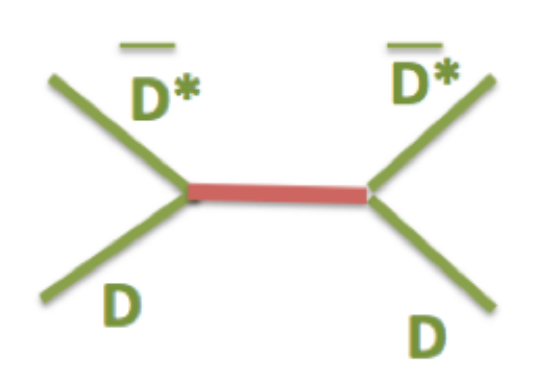}    $\quad$
\includegraphics[width=0.3\textwidth,clip]{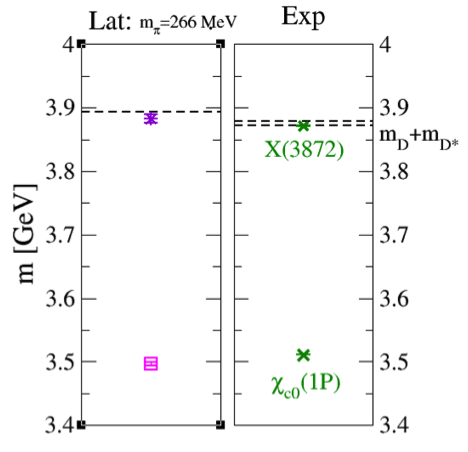}    $\quad$
\includegraphics[width=0.38\textwidth,clip]{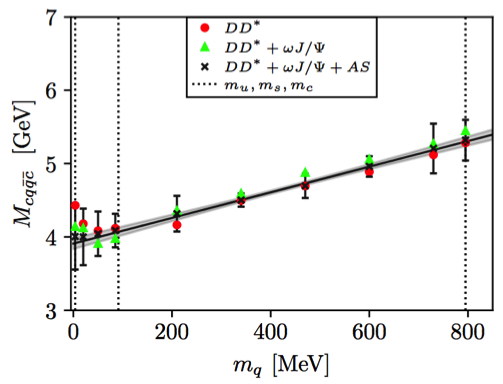}    
\caption{The mass of the bound state $X(3872)$ from lattice \cite{Prelovsek:2013cra} (middle) and the mass of $X(3872)$ as a function of light-quark mass from Dyson-Schwinger approach \cite{Wallbott:2019dng}.}
\label{fig:9}
\end{figure}

The charmonium-like state $X(3872)$ was discovered by Belle \cite{Choi:2003ue}  on the $D\bar D^*$ threshold, therefore any theoretical study needs to take the threshold effect into account. This is done by considering $D\bar D^*$ scattering and looking for poles as described in Section 2.  The only lattice QCD simulations that 
accomplished this \cite{Prelovsek:2013cra,Padmanath:2015era} indeed found the bound state pole related to $X(3872)$ very closely below $D\bar D^*$ threshold (Fig. \ref{fig:9}). It was found that $\bar cc$ and $D\bar D^*$ Fock components are crucial, while diquark-antidiquark is less important  \cite{Padmanath:2015era}. No charged partner was found up to $4.2~$GeV, in agreement with experiment. 

The $D\bar D^*$ channel was recently explored also using a Dyson-Schwinger approach and the position of the pole in the scattering matrix is shown as a function of light-quark mass in Fig. \ref{fig:9} \cite{Wallbott:2019dng}. 
The pole is at $m=3916(74)~$MeV at the physical quark mass, so it could be below or above threshold. This study omitted a $\bar cc$ Fock component and $\bar qq$ annihilation (interestingly, the lattice study has not found a pole in the case of such simplification). \\
 
\section{Theoretical predictions of yet-undiscovered hadrons}

 \underline { Strongly stable doubly-bottom tetraquarks: $\bar b\bar bud$ and $\bar b\bar bud$}
 
\begin{figure}[h!]
\centering
\includegraphics[width=0.2\textwidth,clip]{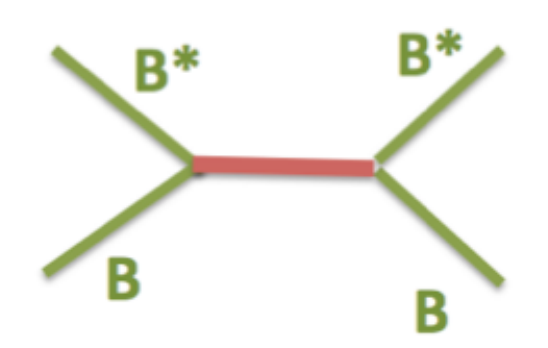}    $\quad$
\includegraphics[width=0.3\textwidth,clip]{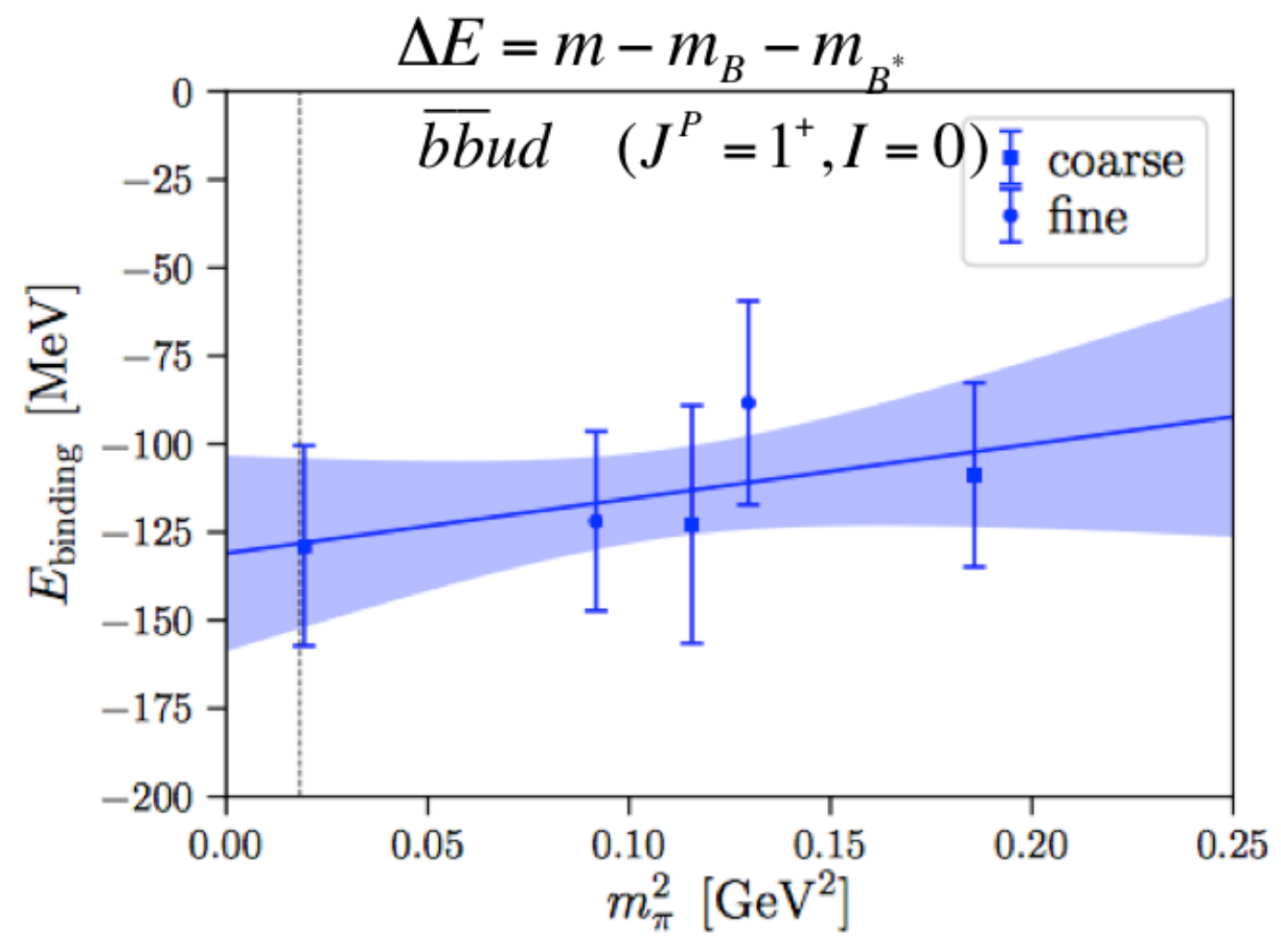}    $\quad$
\includegraphics[width=0.38\textwidth,clip]{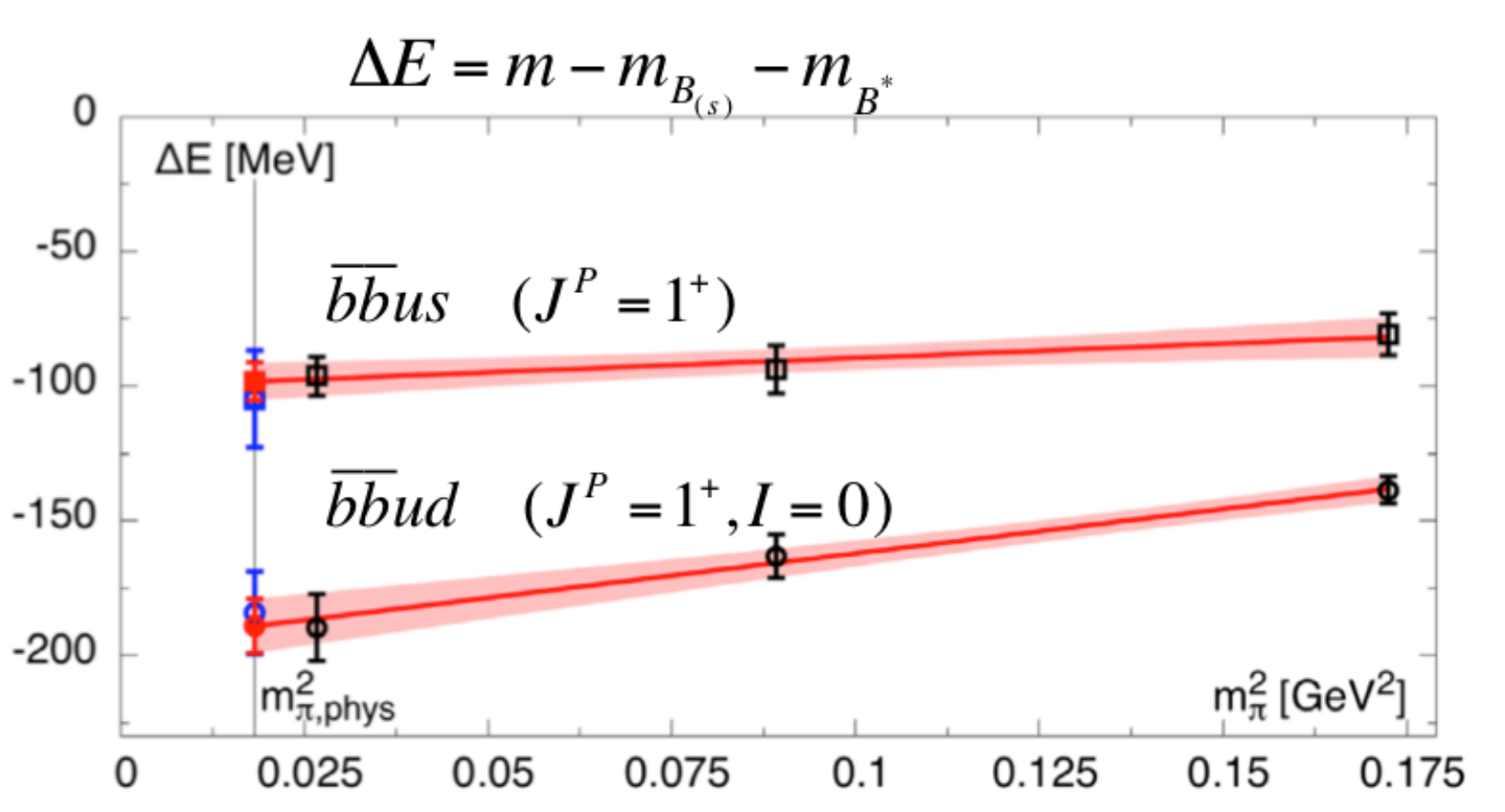}    
\caption{  Lattice prediction for  masses of $\bar b\bar bud$ and $\bar b\bar bus$ with $J^P=1^+$ as a function of $m_\pi$:  \cite{Leskovec:2019ioa} in the middle and \cite{Francis:2016hui} on the right.}
\label{fig:10}
\end{figure}

  Several lattice QCD approaches  \cite{Leskovec:2019ioa,Francis:2016hui,Bicudo:2015vta} agree on the existence of doubly-bottom tetraquarks $\bar b\bar bud$ and $\bar b\bar bud$ with $J^P=1^+$ below strong decay threshold: see Fig. \ref{fig:10} on predicted masses as a function of $m_\pi$. The left plot  was obtained by considering $B B^*$ scattering  in \cite{Leskovec:2019ioa}. The existence of   $\bar b\bar bud$ is implied also by the observed doubly-charmed $\Sigma_{cc}^+$, as argued in \cite{Karliner:2017qjm,Eichten:2017ffp}. Unfortunately, these states will be difficult to discover experimentally.   The consensus on whether the analogous    $\bar c\bar cud$ state is strongly stable has not been reached yet, but it is likely not below threshold. \\

\newpage 

\underline {  Highly excited bottomonia and bottomonium hybrids: $\bar b~b$, $\bar b~glue~b$}

Bottomonia present the richest spectrum of quarkonia below open-flavor threshold. 
The extensive study of highly excited bottomonia has been recently performed by S. Ryan \cite{Ryan:2020spd}   with relativistic $b$ quarks and $m_\pi\simeq 400~$MeV. Each box in Figures on slide 18 of  \cite{ryan_bb_2} represents a predicted bottomonium with a  spin up to four. The states indicated by red represent hybrid candidates $\bar b~glue~b$, where the gluon field is excited. Previous lattice calculation of excited bottomonia was presented in \cite{Wurtz:2015mqa} (and  presented spectra  also for $B$, $B_s$ and $B_c$ excitations). 
 Many of these states, particularly the hybrids, are awaiting experimental discovery. 
 The study of hybrids based on the effective field theory and lattice results was performed in  \cite{Brambilla:2018pyn}. \\

 \underline{Scalar and axial $B_s$}
 
\begin{figure}[h!]
\centering
\includegraphics[width=0.32\textwidth,clip]{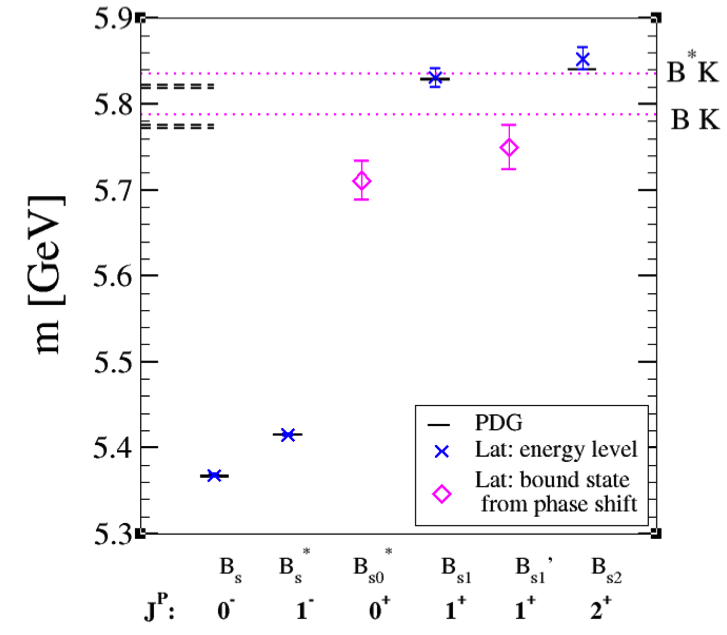}   
\includegraphics[width=0.32\textwidth,clip]{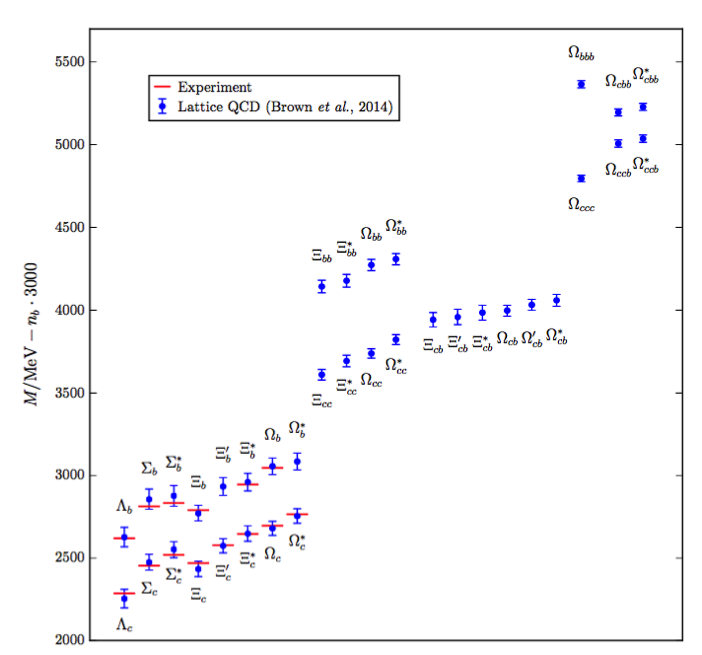}    
\includegraphics[width=0.32\textwidth,clip]{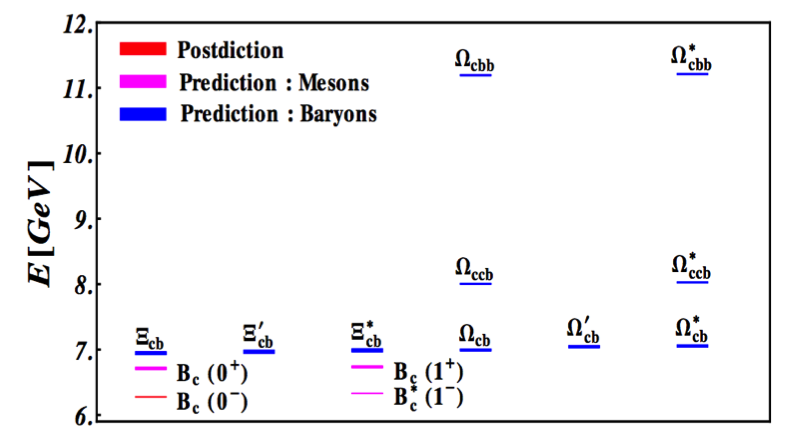}    
\caption{  Lattice QCD predictions for the missing scalar and axial $B_s$ mesons  \cite{Lang:2015hza} (left), baryons with $b$ and/or $c$  \cite{Brown:2014ena} (middle) and hadrons with $b$ and $c$  \cite{Mathur:2018epb} (right).  }
\label{fig:12}
\end{figure}

 In the spectrum of $B_s$ mesons, the scalar and one of the axial mesons are still awaiting  experimental discovery. The lattice prediction for masses of these states in Fig. \ref{fig:12}  (left) took into account the effects of   nearby $B^{(*)} K$ thresholds \cite{Lang:2015hza}. Other observed states agree with masses calculated from lattice.  The charmed partner $D_{s0}(2317)$ of scalar $B_{s0}$ has interesting properties due to closeness to  $DK$ threshold. \\

\underline {  Other hadrons with $b$ and/or $c$ quarks}

There are plenty of hadrons with one or more $b$ and/or $c$ quarks awaiting experimental discovery, as shown by lattice predictions in Fig. \ref{fig:12}. If some of these hadrons are above or near the strong decay threshold, these predictions omit their effects. \\

\underline{Non-existence of strongly stable $\bar bb\bar bb$}

The lattice simulation of $\bar bb\bar bb$ system indicates that there are no strongly stable tetraquarks below corresponding thresholds in channels $J^{PC}=0^{++}, ~1^{++}, 2^{++}$
\cite{Hughes:2017xie}. To reach this conclusion it was essential to take all Wick contractions into account (expect for weak annihilation of $\bar bb$); taking just a part of the Wick contractions leads to  false bound states. This study does not forbid the existence of strongly-decaying resonances. \\

\section{The challenge of coupled-channel scattering}

\begin{figure}[h!]
\centering
\includegraphics[width=0.7\textwidth,clip]{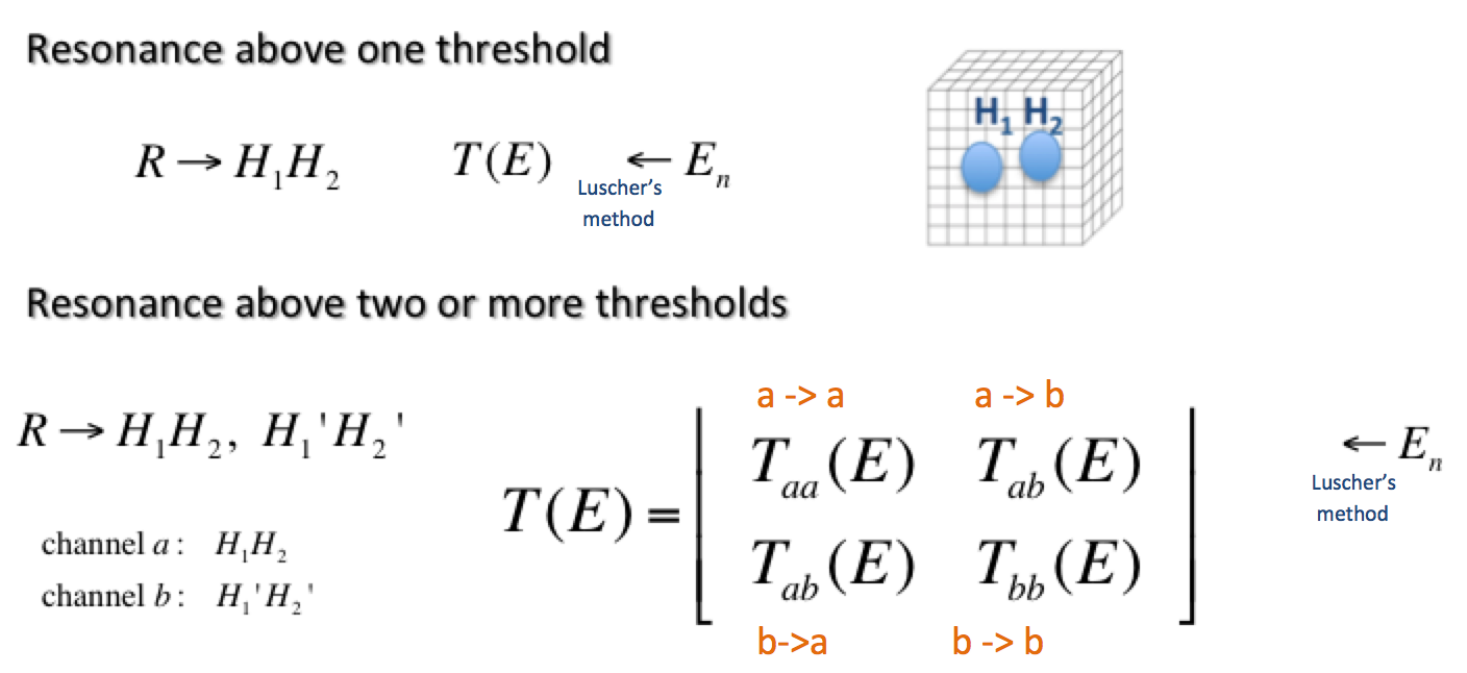}    
\caption{   Strategy of investigating resonance above one or two thresholds on the lattice. The second case requires determination of the coupled-channel scattering matrix $T(E)$.   }
\label{fig:13}
\end{figure}

Many of the interesting resonances,  for example $Z_c(4300)^+$, have not been considered on the lattice since these can strongly decay to several final states. If a resonance $R$  strongly decays via two channels $a$ and $b$, one needs to extract the $2\times 2$ scattering matrix in Fig. \ref{fig:13}. This can be in principle determined from lattice eigen-energies $E_n$ of this coupled-channel system via generalization of L\"uscher's method. But   there are three unknowns $T_{aa}(E_n),~ T_{bb}(E_n), ~T_{ab}(E_n)$ for one given $E_n$ and lattice studies had to resort to parametrizations of $T(E)$. The poles of the resulting scattering matrix give information on mass and the decay width of $R$. This has been performed for several resonances containing only light quarks $u,d,s$ by the Hadron Spectrum Collaboration. 

Only one lattice study extracted coupled-channel scattering matrix via L\"uscher's approach in the system containing heavy quarks and it considered   charmed resonances with $J^P=0^+$ and $I=1/2$ \cite{Moir:2016srx}. The scattering matrix for  the scattering in three coupled channels $D\pi$, $D\eta$ and $D_s\bar K$  was determined.  The analytic re-analysis \cite{Albaladejo:2016lbb,Du:2017zvv,Guo:2018tjx}  has shown that  the  lattice eigen-energies from \cite{Moir:2016srx} are consistent with the presence of two poles in the extracted scattering matrix. The lower pole lies close to $m\simeq 2.1~$GeV, it is dominated by $D\pi$   and is a partner of $DK$ dominated $D_{s0}^*(2317)$  (the later was confirmed by lattice studies  \cite{Mohler:2013rwa,Bali:2017pdv}), while the higher pole   is at   $m\simeq 2.4~$GeV.  This scenario naturally resolves the puzzle why the strange $D_{s0}^*(2317)$ and non-strange $D_0^*(2300)$ states have similar mass, arguing that    they belong to different multipltes.  

Note that most of the exotic hadrons can strongly decay to several final states and rigorous lattice studies of those channels are highly awaited. 


\section{Conclusions and outlook}

Many new exciting results on  the exotic and conventional hadrons with heavy quarks have been recently obtained on the experimental and theoretical   side.  Masses of strongly stable hadrons obtained from lattice QCD agree well with experimental masses. The experimentally  discovered exotic hadrons lie above thresholds and can strongly decay. Lattice QCD has made a significant step in extracting scattering matrices for one-channel scattering, which rendered masses and decay widths of many interesting hadrons that lie near or above threshold, mostly in close agreement with experiment. Analogous steps are now being followed by the Dyson-Schwinger approach.  Many of the exotic hadrons can strongly decay to several final states and rigorous lattice studies of those channels are highly awaited.

\vspace{0.5cm}

\textbf{Acknowledgments} 

\vspace{0.1cm}

 Support by Research Agency ARRS (research core funding No. P1-0035 and No. J1-8137) and DFG grant No. SFB/TRR 55 is   acknowledged.  I  thank Jonathan L. Rosner for careful reading of the manuscript. 


\end{document}